\documentclass[twocolumn,secnumarabic, amssymb, superscriptaddress, nobibnotes, aps, pra]{revtex4-2}

\usepackage{graphicx}% Include figure files
\usepackage{dcolumn}% Align table columns on decimal point
\usepackage{bm}% bold math
\usepackage{amsmath,amssymb,mathtools}
\usepackage{stmaryrd}
\usepackage{times}
\usepackage[euler]{textgreek}
\usepackage[bbgreekl]{mathbbol}
\usepackage{graphicx}
\usepackage{bm,color}
\usepackage{natbib,hyperref}
\usepackage[capitalize]{cleveref}
\usepackage{multirow}

\hypersetup{colorlinks=true,linkcolor=blue,citecolor=blue,urlcolor=blue}

\begin{document}
	
	\title{Room temperature mass sensing based on nonlinear optomechanical dynamics: membrane-in-the-middle versus suspended membrane}
	\author{Jiawei Zheng}
	\affiliation{Fujian Key Laboratory of Light Propagation and Transformation \& Institute of Systems Science, 
		College of Information Science and Engineering, Huaqiao University, Xiamen 361021, China}
	\author{Jinlian Zhang}
	\affiliation{Fujian Key Laboratory of Light Propagation and Transformation \& Institute of Systems Science, 
		College of Information Science and Engineering, Huaqiao University, Xiamen 361021, China}
	\author{Yangzheng Li}
	\affiliation{Department of Rehabilitation Medicine, Sir Run Run Shaw Hospital, School of Medicine, Zhejiang University, Hangzhou 310000, China}
	\author{Luis J. Mart\'inez}
	\affiliation{Center for Quantum Optics and Quantum Information, Universidad Mayor, Camino La Pir\'{a}mide 5750, Huechuraba, Chile}
	\author{Bing He}
	\email{bing.he@umayor.cl}
	\affiliation{Center for Quantum Optics and Quantum Information, Universidad Mayor, Camino La Pir\'{a}mide 5750, Huechuraba, Chile}
	\author{Qing Lin}
	\email{qlin@hqu.edu.cn}
	\affiliation{Fujian Key Laboratory of Light Propagation and Transformation \& Institute of Systems Science, 
		College of Information Science and Engineering, Huaqiao University, Xiamen 361021, China}
	
	\begin{abstract}
		How to weigh something as precise as possible is a constant endeavor for human being, and mass sensing has been essential to scientific research and many other aspects of modern society. In this work, we explore a special approach to mass sensing, which is purely based on the classical nonlinear dynamics of cavity optomechanical systems. We consider two types of systems, the mechanical resonator as a suspended membrane inside optical cavity or as a larger movable membrane that separates the optical cavity into two parts. Under a driving laser field with two tones satisfying a specific frequency condition, both systems enter a special dynamical pattern correlating the mechanical oscillation and the sidebands of oscillatory cavity field. After adding the nano-particle, which has its mass $\delta m$ to be measured, to the mechanical membrane as the detector, the cavity field sidebands will exhibit detectable changes, so that the tiny mass $\delta m$ can be deduced from the measured sideband intensities. For the latter system with a membrane in the middle, one can apply an additional single-tone laser field to magnify the modified sidebands much further, achieving an ultra-high sensitivity $\delta m/m\sim 10^{-11}$ ($m$ is the mass of the membrane), even given a moderate mechanical quality factor. The operation range of the sensors is very wide, covering $7$ or $8$ orders of magnitudes. Moreover, a particular advantage of this type of mass sensors comes from the robustness of the realized dynamical pattern against thermal noise, and it enables such mass sensors to work well at room temperature. 
	\end{abstract}
	
	\maketitle
	\section{Introduction}
	Mass sensing plays an important role in modern science and technology, and the measurement of the mass of tiny particles keeps to be a challenge since the ancient times. Hundreds of years ago, mass spectrometer was invented to find the mass of charged particles. However, not all the particles bring charge, like all types of neutral particles, biomolecules, and others. So far, a standard way of measuring the mass of tiny particles is through a mechanical oscillator with its intrinsic frequency $\omega_m$, which is related to its mass $m$ as $\omega_m=\sqrt{k/m}$ ($k$ is the spring constant). If some nanoparticles with the mass $\delta m$ are attached to the mechanical oscillator, this frequency will be modified by the amount
	\begin{align}
	\delta\omega_m/\omega_m=-\delta m/2m.
	\end{align} 
	Then, the mass sensing for the nanoparticles can be converted to measuring the frequency shift $\delta\omega_m$ of the mechanical oscillator. 
	
	To achieve high precision detection, one can reduce the mass $m$ of the mechanical resonator (the detector) to as small as possible, so that a tiny mass $\delta m$ can be detected with a relatively large mechanical frequency shift ratio $\delta\omega_m/\omega_m$. Nowadays, based on this idea, many low-dimensional nanostructures, such as carbon nanotubes, nanowires, and others, e.g., \cite{ms1,ms2,ms3,ms4,ms5,ms6,ms7,ms8,ms9,ms10,ms11,ms12,ms13,ms14,ms15}), have been used to achieve the highly sensitive mass sensing to the levels of sub-zg \cite{ms1} and even yg \cite{ms13}, by the direct electrical measurements of mechanical frequency shift $\delta\omega_m$. Since the thermal noise induces random motion of the mechanical resonator, which will definitely impair the detection 
	precision for mechanical frequency, those implementations of mass sensing must be in a low temperature environments, e.g., as low as $T=4.3$ K in Ref. \cite{ms13}. Obviously, the requirement of cryogenic temperatures limits their applicability. To relax this restriction, one can apply the mechanical modes of microcavities, such as microtoroid cavity, microsphere cavity, etc. (see Refs. \cite{msc1,msc2,msc3,msc4,msc5,msc6,msc7,msc8,msc9,msc10}), as the mechanical resonators, and 
	the associated mechanical frequency shift can be detected through the coupled optical field of these cavity optomechanical systems (COMS). The measurement to the levels of sub-pg \cite{msc3} 
	and $10^{-19}g$ \cite{msc8} can be achieved by such mass sensors.

    An alternative approach to detect a tiny $\delta m$ is to improve the capability of measuring the relative mechanical frequency shift $\delta\omega_m/\omega_m$, which could be as small as possible. 
	Due to the inevitable mechanical damping, however, the measurement of the mechanical frequency shift lower than the mechanical damping rate, i.e., $|\delta\omega_m|<\gamma_m$, is hard for most of the 
	schemes based on a direct measurement of the mechanical frequency. Though one can measure the mechanical frequency shift beyond this limit simply by increasing the detection time, the detection efficiency will be lowered considerably. Recently, some approaches were developed to realize the mass sensing beyond this limit (see, e.g. Refs. \cite{ms14,msb1,msb2,msb3,msb4,msb5}). 
	The sensitivity $|\delta\omega_m|/\gamma_m\sim0.01$ is achievable by the applications of optomechanically induced transparency \cite{msb1}, high-order
	sidebands \cite{ms14,msb2}, quantum criticality \cite{msb5}, or the detection of cavity quadrature \cite{msc7}. However, ultra-low temperatures are also required for those schemes; otherwise the sensitivity will be decreased quickly by thermal noise. More recently, a theoretical scheme considering a suspended membrane (SM) in COMS indicates that an ultra-high sensitivity $|\delta\omega_m|/\gamma_m=10^{-7}\sim10^{-5}$ is achievable \cite{qlin1}. This scheme based on a kind of optomechanical nonlinearity works well at room temperature, since the effect of thermal noise can be well suppressed by an oscillation locking effect. Here, we present an improvement on that scheme by considering a membrane-in-the-middle (MIM) COMS, so that the sensitivity can be improved by one or even two orders ($|\delta\omega_m|/\gamma_m=10^{-9}\sim10^{-7}$) with the realizable system parameters. 
	
	The rest of the paper is organized as follows. Sec. \ref{sec2} is dedicated to explore the mechanism of mass sensing based on different setups of nonlinear optomechanical dynamics. After that, in Sec. \ref{sec3}, the performance of one type of mass sensing is discussed in details. The robustness of the mass sensing against thermal noise is illustrated in Sec. \ref{sec4}, before final section 
	of the discussion and conclusion.

\begin{figure}[t]
	\centering\includegraphics[width=8.5cm]{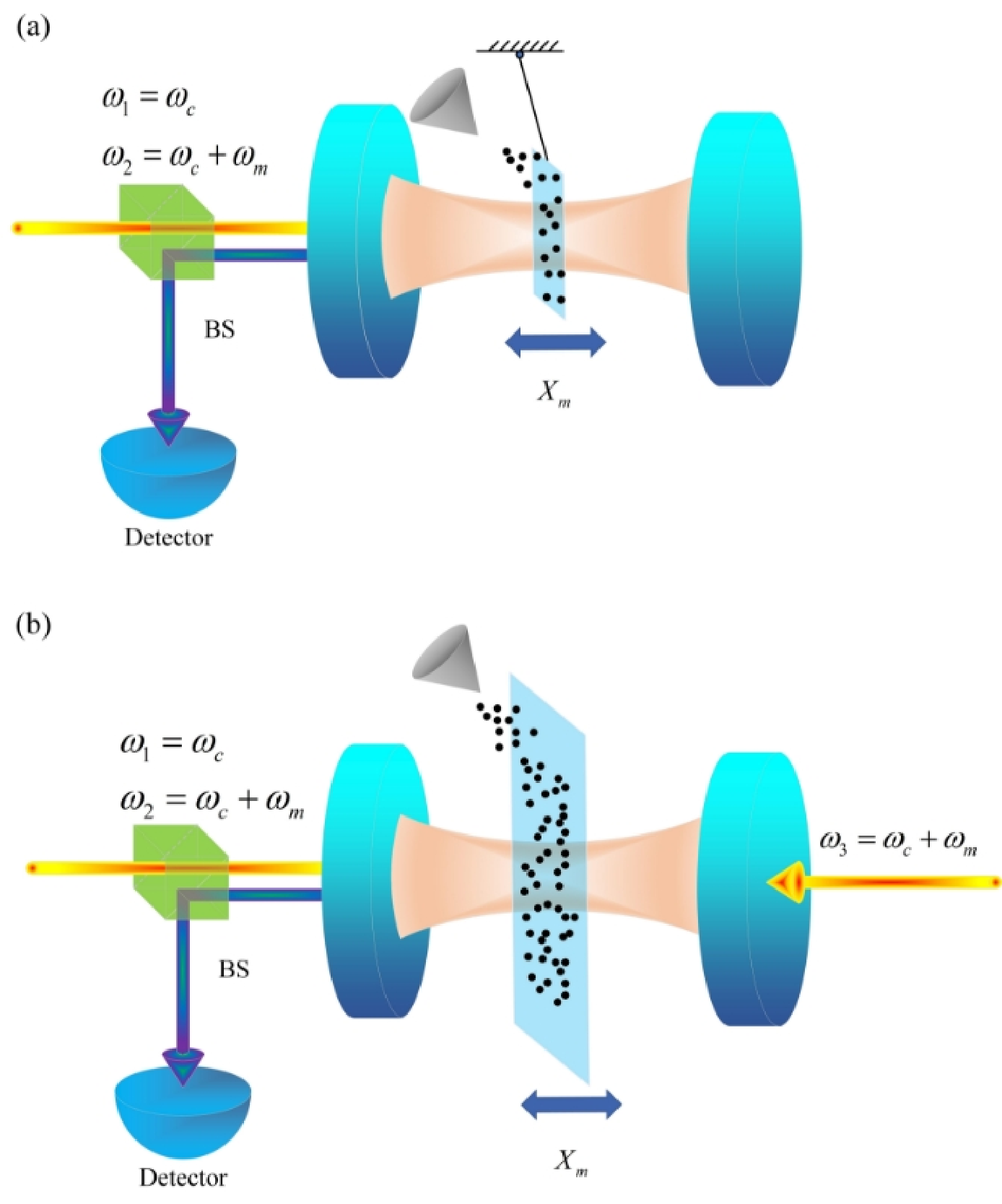}
	\caption{Setups for the mass sensing. Here we consider a system of suspended membrane (SM) \cite{sm1,sm2,sm3} in (a) and a system of membrane in the middle (MIM) \cite{mi1,mi2,mi3,mi4,mi5,mi6,mi7,mi8,mi9} in (b). A two-tone laser field is used as the pump, and a small thin membrane is used as a mechanical resonator. Through the detection of the output field from the cavity, the mechanical frequency shift induced by the attached nanoparticles can be measured. In 
		(b), a large thin membrane is used as a mechanical resonator, which separates the cavity into two parts. At the same time, a blue-detuned laser is injected from the right.}  \label{fig1}
\end{figure}

\begin{figure}[b]
	\centering\includegraphics[width=8.5cm]{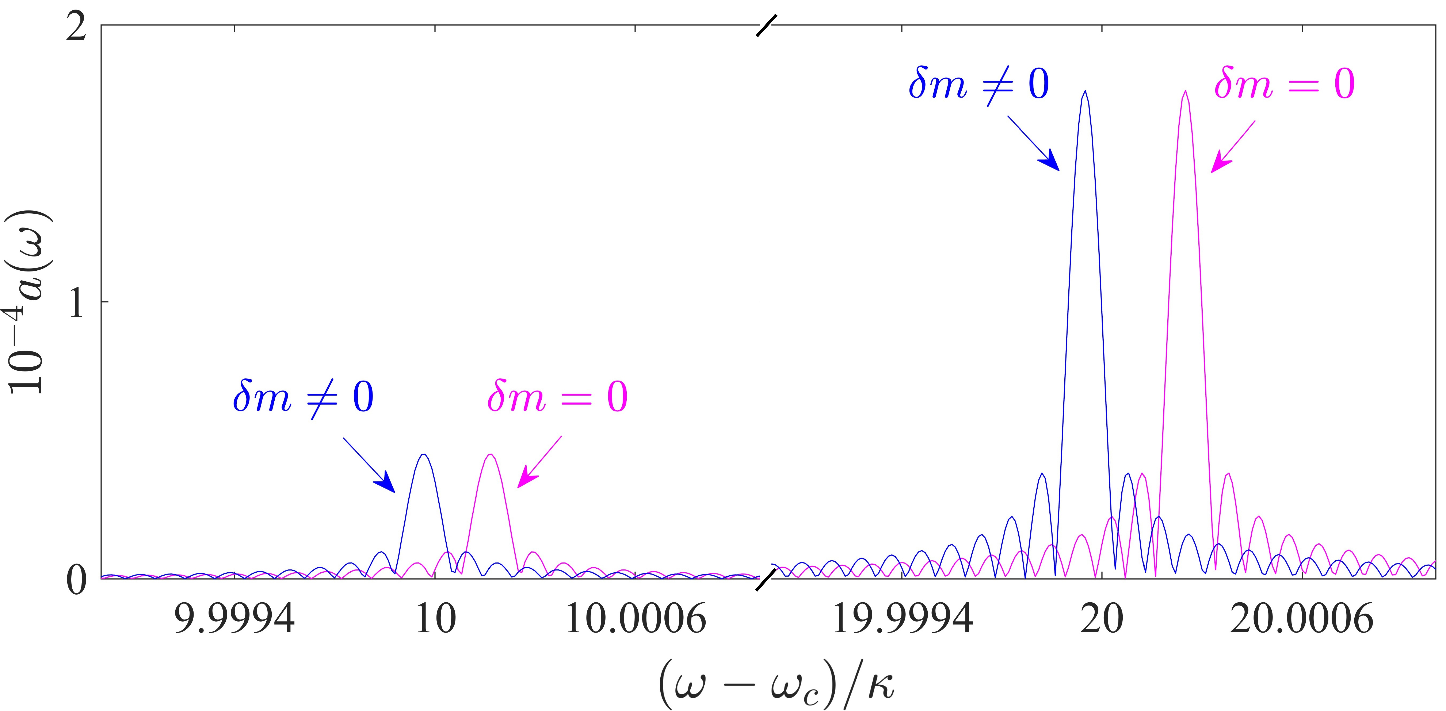}
	\caption{The first-order and second-order cavity field sidebands of a COMS with suspended membrane, when it is driven by a single-tone laser field. The original positions of the sidebands are shifted from $n\omega_m$ ($n$ are integers) due to optical spring effect, and they are shifted further by the variation $\delta\omega_m=-10^{-4}\kappa$ of the theoretical value $\omega_m=\sqrt{k/m}$ after a tiny mass $\delta m$ is added. There are little changes of the sideband amplitude. The system parameters are scaled wit the cavity field damping rate as $g_m=10^{-5}\kappa$, $\omega_m=10\kappa$, $Q=\omega_m/\gamma_m=10^5$, and the single-tone drive with $\Delta=-\omega_m$ has $E=\sqrt{3}\times10^5\kappa$.}  \label{fig2}
\end{figure}

	\section{Setups, dynamics, and mechanism of sensors}
	\label{sec2}
	
	Any COMS has a mechanical mode, such as a movable membrane under radiation pressure; see Fig. 1. The coupling between the mechanical mode and the cavity field mode $a$ through radiation pressure can be formulated as an interaction potential proportional to the coupling constant $g_m$, such that the dynamical equations of the system (in a reference system rotating at the resonant cavity frequency $\omega_c$) can be written as follows:
	\begin{align}
		\dot{a}&=-\kappa a+ig_mX_ma+Ee^{i\Delta t}+\sqrt{2\kappa_i}\xi_c(t),\nonumber\\
		\dot{X}_m&=\omega_mP_m,\nonumber\\
		\dot{P}_m&=-\omega_mX_m-\gamma_mP_m+g_m|a|^2+\sqrt{2\gamma_m}\xi_m(t),
		\label{dy}
	\end{align}
	where $X_m$ and $P_m$ are the dimensionless displacement and momentum of the mechanical membrane, respectively, and the pump field detuning is defined as $\Delta=\omega_c-\omega_{L}$. Here, the driving field amplitude is related to the pump power $P$ as $E=\sqrt{2\kappa_e P/\hbar\omega_L}$, where the cavity damping rate $\kappa=\kappa_e+\kappa_i$ consists of the coupling rate $\kappa_e$ to the cavity and the intrinsic loss rate $\kappa_i$. During the mass sensing operation, the mechanical membrane is in an thermal environment, which drives the membrane by stochastic force $\sqrt{2\gamma_m}\xi_m(t)$ in terms of the mechanical damping rate $\gamma_m$ and with the noise term satisfying the correlation $\langle \xi_m(t)\xi_m(t')\rangle=(2n_{th}+1)\delta(t-t')$ 
	($n_{th}=1/(e^{\hbar \omega_m/k_BT}-1)$) \cite{noise}. On the other hand, the cavity noise $\sqrt{2\kappa_i}\xi_c(t)$ acting from another environment with the effective zero temperature ($n_{th}=0$) has a negligible effect on the concerned processes.
	
	The mechanical membrane will be in oscillation, which was called self-sustained or back-action-induced oscillations, once the pump laser power is over a threshold of the Hopf bifurcation (see, e.g. \cite{ss1,ss2,ss3,ss4,OMS,ss5,ss6,ss7}). 
	Under a single-tone pumping laser field with a detuning $\Delta$, the mechanical oscillation frequency will always be shifted by a certain amount $\delta$ due to optical spring effect \cite{OMS}, and it can even significantly determine the dynamical evolution of the system \cite{coms}, so the oscillation of mechanical membrane takes the form
	\begin{align}
		X_m=A_m\cos(\Omega_m t)+d_m,
	\end{align}
where $\Omega_m=\omega_m+\delta$, $A_m$ the amplitude of the oscillation, and $d_m$ a static displacement.
	After the addition of the mass $\delta m$ of nanoparticles to the mechanical membrane, there will be a further shift $\delta\omega_m$ of the mechanical frequency, as well as a slight amplitude change $\delta A_m$ to the mechanical amplitude at the same time, so that $A'_m=A_m+\delta A_m$. In this situation, the stabilized oscillatory cavity field takes the form
	\begin{align}
		a(t)= e^{i\phi(t)}\sum_l\alpha_{l}e^{i(l\cdot\Omega_m+\Delta) t},
		\label{fs1}
	\end{align}
	where $l$ are the integers, $\phi(t)=g_mA'_m/\Omega_m\sin(\Omega_mt)$, and
	\begin{align}
		\alpha_{l}&=\frac{E}{\kappa}\frac{J_l(-g_mA'_m/\Omega_m)}{il\cdot\Omega_m/\kappa+1-i(g_md_m-\Delta)/\kappa},\nonumber\\
		\Omega_m &=\omega_m+\delta+\delta \omega_m,
		\label{field}
	\end{align}
	 where $J_l$ is the $l$th-order Bessel function of the first kind. In Fig. 2, we compare a couple of cavity field sidebands before and after adding the mass $\delta m$, with a COMS of $\omega_m=10\kappa$.
	Each pair of the sidebands due to a pump field with $\Delta=-\omega_m=10\kappa$ displays a shift $\delta\omega_m$ of the sideband frequencies, since the distance between each sideband is equal to the oscillation frequency of the mechanical membrane. Then, the added mass $\delta m$ can be detected through this frequency shift $\delta\omega_m$ of the cavity sidebands. At the present time, Pound-Drever-Hall (PDH) technique \cite{PDH1,PDH2,PDH3} can resolve a frequency shift up to $\delta\omega_m\sim10^{-3}\kappa$ through the spectrum, but a higher resolution for the even smaller $\delta\omega_m$ is hard to achieve. 
	
		\begin{figure}[tb]
		\centering\includegraphics[width=8.5cm]{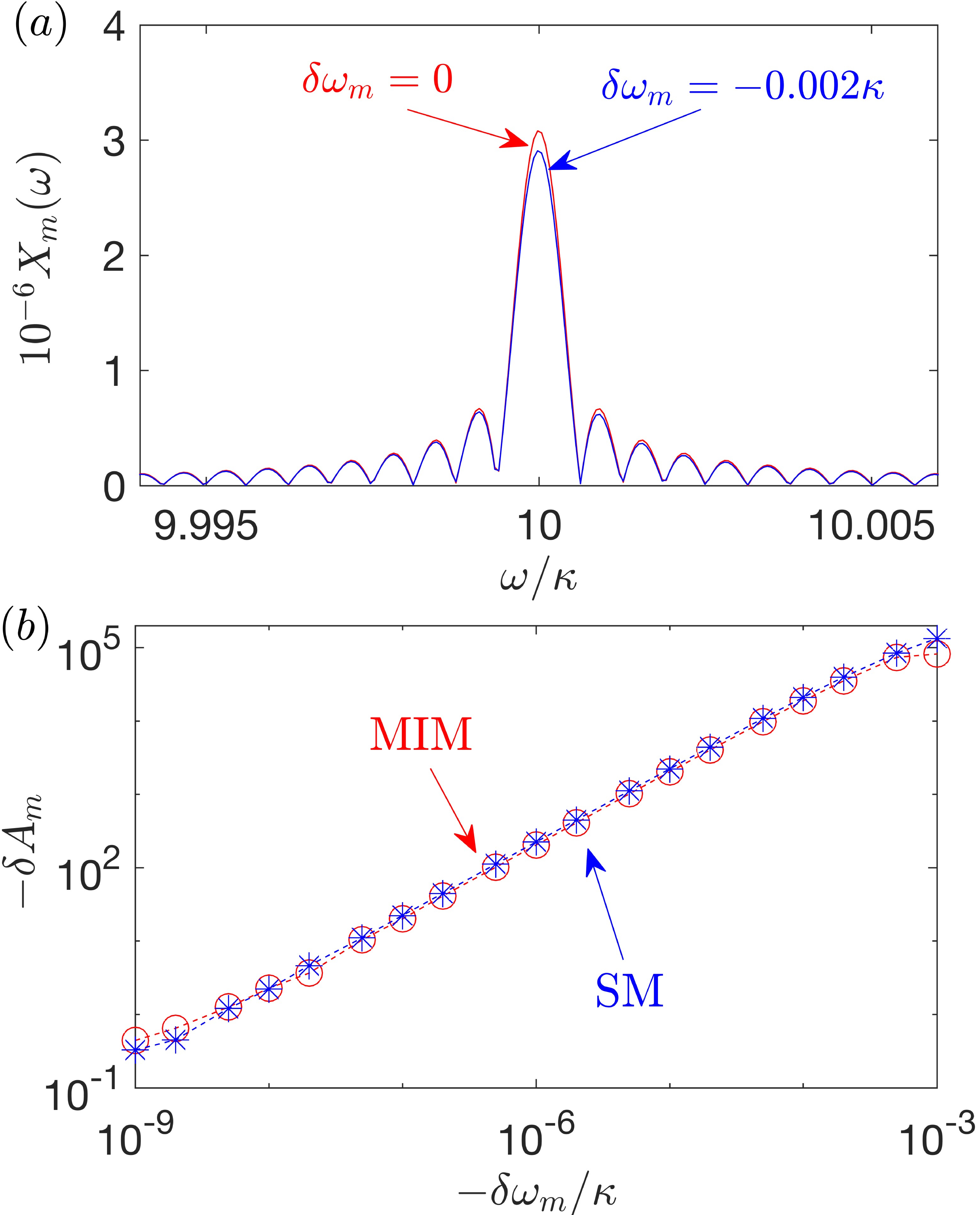}
		\caption{(a) The stabilized mechanical spectrum of the system with SM. This actually observed mechanical oscillation frequency is locked even with a small shift $\delta \omega_m$ from the theoretical value $\omega_m=\sqrt{k/m}$, but the amplitude of the mechanical oscillation will be modified. (b) A scaling relation between the observed mechanical amplitude modification and the small variation of the theoretical mechanical frequency $\omega_m=\sqrt{k/m}$. Within the displayed range, the linear relation $\delta A_m=const\cdot \delta\omega_m$ is exact. Here we adopt the logarithmic scales for both horizontal and vertical axis. The drive amplitude is $E=\sqrt{3/2}\times10^5\kappa$ for each tone on the SM system and $E=10^5\kappa$ for each tone on the MIM system, and the system parameters are the same as those in Fig. \ref{fig2}.}  \label{dam}
	\end{figure}

	In this work, we will consider the application of driving fields with multiple tones. For example, the pump carries two frequencies so that the first equation of Eq. (\ref{dy}) will be replaced by 
		\begin{align}
		\dot{a}&=-\kappa a+ig_mX_ma+\sum_{i=1}^2E_ie^{i\Delta_i t}.
		\label{dy2}
	\end{align}
	One such choice is a two-tone pumping laser field with a fixed frequency difference of its frequency components $\omega_{L_1}$ and $\omega_{L_2}$ being the same as the original mechanical frequency $\omega_m$, i.e., satisfying the frequency match condition:
	\begin{align}
		|\omega_{L_1}-\omega_{L_2}|=\omega_m,
		\label{match}
	\end{align} 
	where $\omega_m=\sqrt{k/m}$ is the theoretical mechanical frequency determined by the system fabrication. Under this condition, the threshold for entering optomechanical oscillation can be significantly reduced \cite{ns,ofc}. Most importantly, there will exist a total locking of the three elements of the realized mechanical oscillation---its amplitude and frequency, as well as the oscillation phase, will be locked irrespective of the pump power that can be adjusted over a considerable range \cite{level1,level2}.

 Due to the mechanism of oscillation locking, the actually observed mechanical oscillation frequency will be still locked to the difference $|\Delta_1-\Delta_2|$ of the two drive tones, which can be adjusted as close to the theoretical frequency $\omega_m$ as possible, if this theoretical mechanical frequency $\omega_m$ is only modified within a small range. After a tiny mass $\delta m$ is attached to the mechanical membrane, the theoretical mechanical frequency $\omega_m=\sqrt{k/m}$ will be varied by a small amount according to Eq. (1), e.g., $\delta\omega_m=-2\times10^{-3}\kappa$ in Fig. \ref{dam}(a). However, the observed oscillation frequency is still locked to the original $\omega_m$, given two pump tones exactly satisfying Eq. (\ref{match}). Such a suppression of optical spring effect ($\delta=0$) and the immunity to tiny deviations $\delta \omega_m$ lead to the following form of the stabilized mechanical motion:
	\begin{align}
		X_m=A'_m\cos(\omega_mt)+d_m. \label{mm}
	\end{align}
	Because the mechanical frequency is thus locked, the effect of added mass is now manifested by a modified mechanical amplitude to $A'_m=A_m+\delta A_m$; see Fig. \ref{dam}(a). An interesting feature is that, within a certain range of $\delta \omega_m$, the modification $\delta A_m$ is linearly proportional to $\delta \omega_m$, as illustrated in Fig. \ref{dam}(b) showing such scaling for both SM system and MIM system. This linear scaling range is important to our mass sensing operation.
	
	The corresponding peak positions of the cavity field sidebands are at $\omega_c+n\omega_m$ ($n$ is integer). Under a two-tone drive acting on the system of SM and satisfying Eq. (\ref{match}), the cavity field $a(t)$ takes the following approximate form in the reference system rotating at the frequency $\omega_c$:	
	\begin{align}
		a(t)&\approx e^{i\phi(t)}\sum_l\left[\alpha_{l,1}e^{i(l\cdot\omega_m+\Delta_1) t}+\alpha_{l,2}e^{i(l\cdot\omega_m+\Delta_2) t}\right],\nonumber\\
		&=e^{i\phi(t)}\sum_l(\alpha_{l,1}+\alpha_{l+1,2})e^{i\omega_l t}e^{i\Delta_1 t},
		\label{fs}
	\end{align}
	where $l$ is an integer, and $\phi(t)=g_mA'_m/\omega_m\sin(\omega_mt)$. The involved sidebands have 
	\begin{align}
		\alpha_{l,1(2)}&=\frac{E}{\kappa}\frac{J_l(-g_mA'_m/\omega_m)}{il\cdot\omega_m/\kappa+1-i(g_md_m-\Delta_{1(2)})/\kappa},
		\label{field}
	\end{align}
	with $\omega_l=l\cdot\omega_m$ in the reference frame rotating at the frequency $\omega_c$ (it is differed from the actual frequency by $\omega_c$). Through measuring the modification of the higher-order sideband amplitudes, $\alpha_{l,1}+\alpha_{l+1,2}$ ($l\geq 2$), the mass sensing with $\delta\omega_m/\gamma_m\sim10^{-5}$ can be achieved \cite{qlin1}.

	\begin{figure}[b]
	\centering\includegraphics[width=8cm]{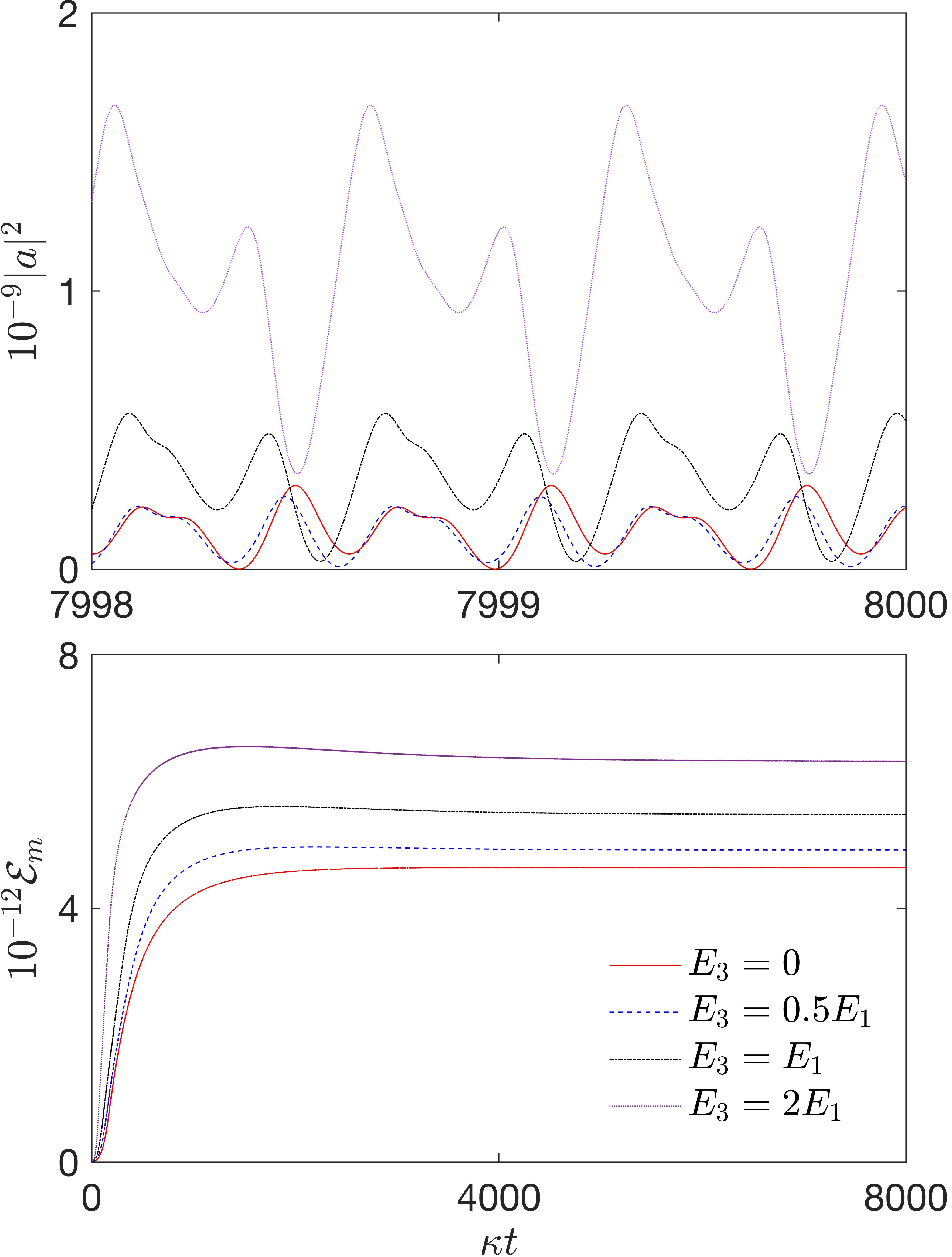}
	\caption{An illustration of the effect of the third pump tone acting on a MIM system. Here, the two tones satisfying Eq. (\ref{match}) have $E_2=E_1=10^5\kappa$, while the third tone is gradually increased from zero. The other fixed system parameters are the same as those in Fig. \ref{fig2}.}  \label{fige}
\end{figure}

	A step towards improving the mass sensing much further is to consider a system of MIM in Fig. 1(b), where an extra pump is also introduced on the other side of the membrane. Suppose the extra pump field is a blue-detuned one with $\Delta_3=-\omega_m$, the dynamical equations of the system will be modified to
    \begin{align}
			\dot{a}_1&=-\kappa a_1-iJa_2+ig_mX_ma_1
			+\sum_{i=1}^2E_ie^{i\Delta_i t},\nonumber\\
			\dot{X}_m&=\omega_mP_m,\nonumber\\
			\dot{P}_m&=-\omega_mX_m-\gamma_mP_m+g_m(|a_1|^2-|a_2|^2)\nonumber\\
			&+\sqrt{2\gamma_m}\xi_m(t),\nonumber\\
			\dot{a}_2&=-\kappa a_2+iJa_1-ig_mX_ma_2+E_3e^{-i\omega_m t}.
			\label{dy3}
		\end{align}
	The separation of the optical cavity by the membrane in the middle introduces one more system parameter $J$ as the coupling strength of two cavity modes. How strong the coupling between two cavity modes will only modify the cavity field sideband amplitudes slightly (it is why we will not discuss about how the figures-of-merit change with the coupling strength $J$ below), because the exchange of photons is mutual.
	However, the introduction of the extra degrees of freedom and an extra pump field will considerably alter the system's response to the modified theoretical mechanical frequency by $\delta \omega_m$ and thus impact on the concerned mass sensing performance. 
	
		\begin{figure}[b]
		\centering\includegraphics[width=8.5cm]{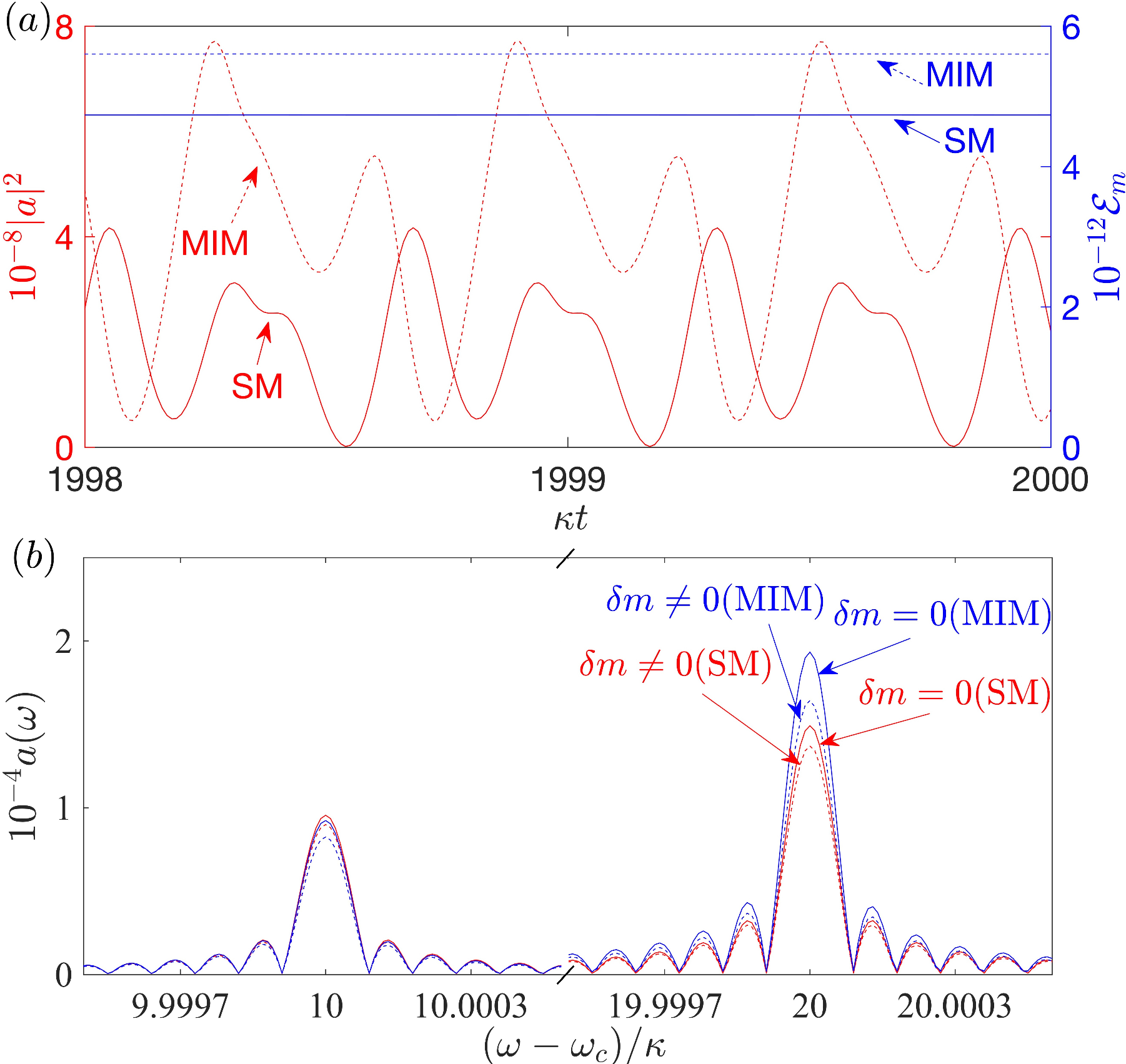}
		\caption{(a) The comparison of the real-time mechanical energy $\mathcal{E}_m$, as well as the cavity photon numbers $|a|^2$, between a COMS of SM and a corresponding one of MIM. (b) The corresponding first-order and second-order cavity field sidebands. The positions of them are locked by two drive tones satisfying Eq. (\ref{match}), in spite of a mechanical frequency shift 
			$\delta\omega_m=-10^{-4}\kappa$ induced by a small mass $\delta m$, but the sideband amplitudes (especially that of the the second-order one) will be obviously modified by the $\delta m$. We choose the drive amplitude $E=\sqrt{3/2}\times10^5\kappa$ for the SM system and $E=10^5\kappa$ for the MIM system, respectively, so that the totally consumed powers are the same. For the MIM system, we assume $J=\kappa$, and the other fixed parameters are the same as those in Fig. \ref{fig2}.}  \label{fig22}
	\end{figure}
	
	There are two main features of dynamics for this type of MIM systems. One of them is the same as with the system of SM in Fig. 1(a): the stabilized mechanical oscillation frequency will be locked by the two-tone field pumping from the left side if the deviation $\delta \omega_m$ induced by the added mass is sufficiently small. Without loss of generality, we assume here that the actually observed mechanical frequency will be equal to $\omega_m$ under the setting $|\Delta_{1}-\Delta_{2}|=\omega_m$ for the two drive tones. Certainly, if the frequency difference of the two tones is close to but is not exactly equal to $\omega_m$, the actually stabilized mechanical oscillation frequency will be locked to this frequency difference too. Then the solution of the stabilized cavity field $a_1(t)$ in the left cavity can be in the same form as in Eq. (\ref{fs}), and the corresponding peak positions of the field sidebands are as well fixed to $\omega_c+n\omega_m$ ($n$ is an integers). On the other hand, the pump field injected to the right cavity will push the membrane from the opposite direction and modify the mechanical amplitude, which will in turn alter the sideband amplitudes of the cavity fields. This effect is well described in Fig. \ref{fige}, which displays how a gradually increased third pump changes the stabilized cavity field photon number and the associated mechanical energy. With the action of the third pump, therefore, it is possible to magnify the modification of the cavity field sidebands caused by the tiny mass added to the membrane. As a study on the general principle of a mass sensing, we will only choose two drive amplitudes of the third pump, i.e., $E_3=E$ and $E_3=2E$ ($E_1=E_2=E$), as the examples for illustrating the advantages of MIM systems.

We present a detailed comparison between the dynamical processes of a SM system and an MIM system in Fig. \ref{fig22}. For this comparison we adopt the same total pump power to the different systems, so that each tone of the pump field on the SM system is higher (the sum of the powers of the two tones is equal to the corresponding sum of the three drive tones for the MIM system). Even with the same total pump power, the responses of the different systems are very different as shown in Fig. \ref{fig22}(a). The system of MIM will stabilize in a state with much more mechanical energy and cavity photons, so the corresponding field sideband amplitudes will be larger than those of the SM system. From Fig. \ref{fig22}(b) one sees that the modification of the second-order sideband by a small mass, which induces a frequency deviation 
$\delta\omega_m=-10^{-4}\kappa$, is much more obvious to the system of MIM. This example of quantitative comparison explains why the more complicated systems of MIM worth a consideration for mass sensing.

\begin{figure}[b]
	\centering\includegraphics[width=8.5cm]{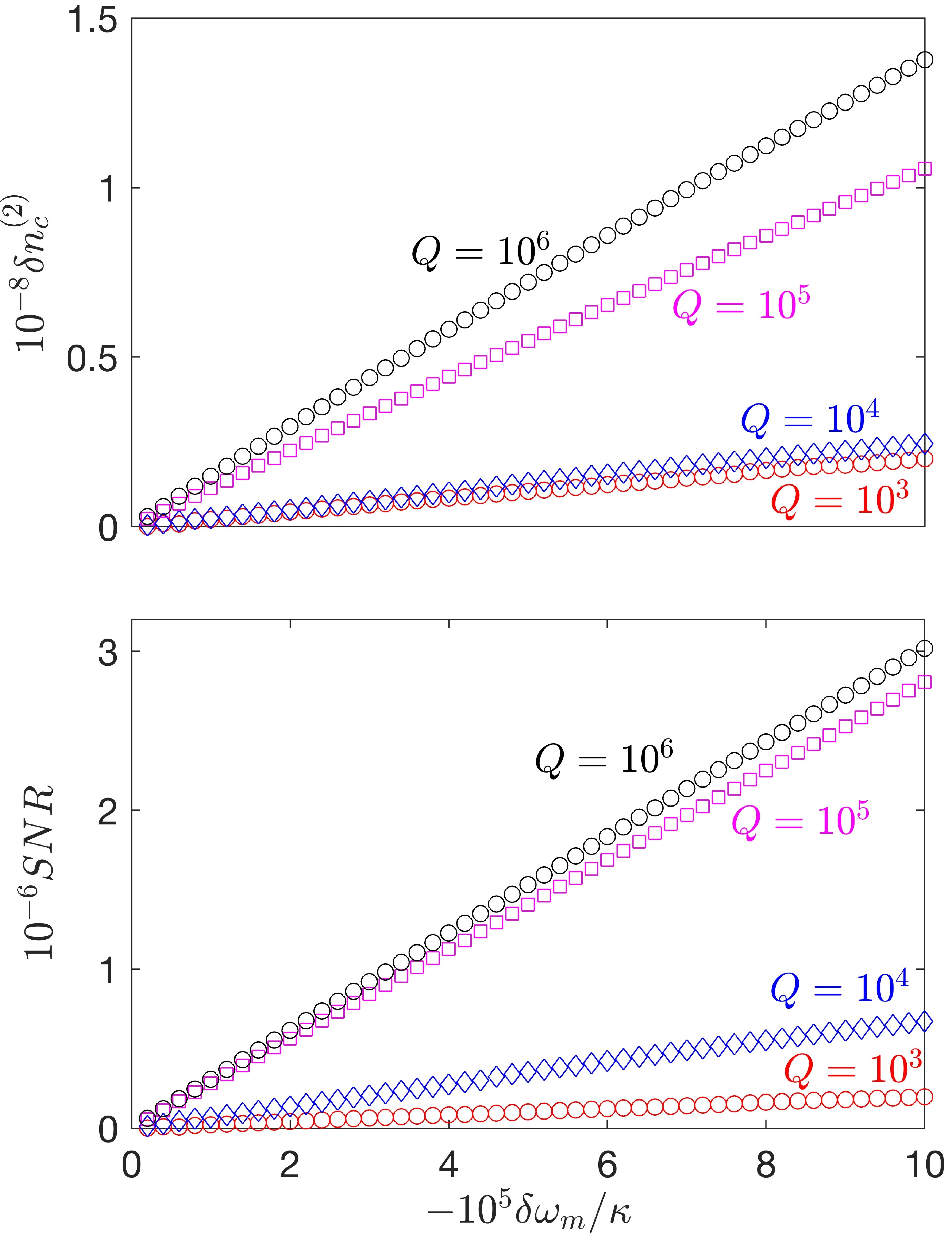}
	\caption{Linear scaling relations between the modifications of the second-order sideband intensities and the mechanical frequency shifts $\delta \omega_m$ induced by the added masses, as well as the corresponding scaling relations between the defined SNR and $\delta \omega_m$. Four different mechanical quality factors $Q=\omega_m/\gamma_m$ are considered here, for the mechanical frequency shifts from $\delta\omega_m=-10^{-6}\kappa$ to $\delta\omega_m=-10^{-4}\kappa$. The fixed system parameters are the same as those in Fig. \ref{fig2}, and the drive tone amplitudes are the same as the one in Fig. \ref{dam}.}  \label{fig3}
\end{figure}	
	
\section{Mass sensing performance}
	\label{sec3}
	
	From now on, we mainly consider the systems of MIM and study their figures-of-merit in mass sensing. We choose the second-order sideband at the frequency $\omega_2=2\omega_m+\omega_c$ for the purpose. 
	The drive detunings are set as $\Delta_1=0$, $\Delta_2=-\omega_m$ and $\Delta_3=-\omega_m$. The intensity of the second-order sideband, 
	\begin{align}
		n^{(2)}_c(\omega_2)=|\alpha_{2,1}+\alpha_{3,2}|^2,
	\end{align}
	is to be measured in the operation.
	According to the input-output relation,
	\begin{align}
		a_{out}(t)=&a_{in}(t)-\sqrt{\kappa_e}a(t),
	\end{align}
	the corresponding output intensity of this second-order sideband is $n^{(2)}_{c,out}(\omega_2)=\kappa_e n^{(2)}_c(\omega_2)$, since the pump field does not have the frequency component at $\omega_2$.
Here, without loss generality, we set a detection time of $t_d=1$ second in the measurement of $n^{(2)}_{c,out}(\omega_2)$.

In reality, the operation of a measurement is limited by the shot noise, which arises from a statistics property and is proportional to the square root of the measured intensity. To keep a lower standard deviation in the measurement, the 
detection process should be repeated for many times and the result is their average. However, a long time for measurement will certainly lower the mass sensing efficiency. 
To indicate such efficiency of the measurement, we therefore adopt the following signal-to-noise ratio (SNR):
	\begin{align}
		\text{SNR}&=\frac{|n^{(2)}_{c,out}(\omega_2;0)-n^{(2)}_{c,out}(\omega_2;\delta\omega_m)|}{\sqrt{ \kappa n^{(2)}_{c,out}(\omega_2;0)}+\sqrt{ \kappa n^{(2)}_{c,out}(\omega_2;\delta\omega_m)}}\nonumber\\
		&=\left|\sqrt{ n^{(2)}_{c,out}(\omega_2;\delta\omega_m)/\kappa}-\sqrt{ n^{(2)}_{c,out}(\omega_2;0)/\kappa}\right|,\label{snr}
	\end{align}
where $n^{(2)}_{c,out}(\omega_2;0)$ and $n^{(2)}_{c,out}(\omega_2;\delta\omega_m)$ denote the output intensity of the second-order sideband before and after adding the mass $\delta m$, respectively.

In a typical range of the frequency shift $\delta\omega_m=-10^{-6}\kappa\sim-10^{-4}\kappa$, we consider the change $\delta n_c^{(2)}=|n^{(2)}_{c}(\omega_2;0)-n^{(2)}_{c}(\omega_2;\delta\omega_m)|$, which should be linear with $\delta \omega_m$ after considering the first-order Taylor expansion, $n^{(2)}_{c}(\omega_2;\delta\omega_m)\approx n^{(2)}_{c}(\omega_2;0)+\frac{dn^{(2)}_{c}}{d(\delta\omega_m)}\delta\omega_m$. 
The exact numerical calculations well prove this relation demonstrated in Fig. \ref{fig3}(a). Correspondingly, such difference can be translated to the plots of SNR in Fig. \ref{fig3}(b), which are also linearly proportional to the shifted theoretical mechanical frequency $\delta\omega_m$ by the tiny mass $\delta m$ to be measured.
A larger mechanical quality factor $Q=\omega_m/\gamma_m$ obviously improves the SNR in Fig. \ref{fig3}. Here we fix the theoretical mechanical frequency $\omega_m$, so a higher mechanical factor means a lower mechanical damping rate $\gamma_m$. Under this assumption, the mechanical amplitude change, as well as the corresponding output sideband change, will be higher with a lower mechanical damping rate. The quality factor $Q=10^5$ we adopt in Fig. \ref{fig3} is an upper bound that can be reached in a usual experiment. 

\begin{figure}[h]
	\centering\includegraphics[width=8.5cm]{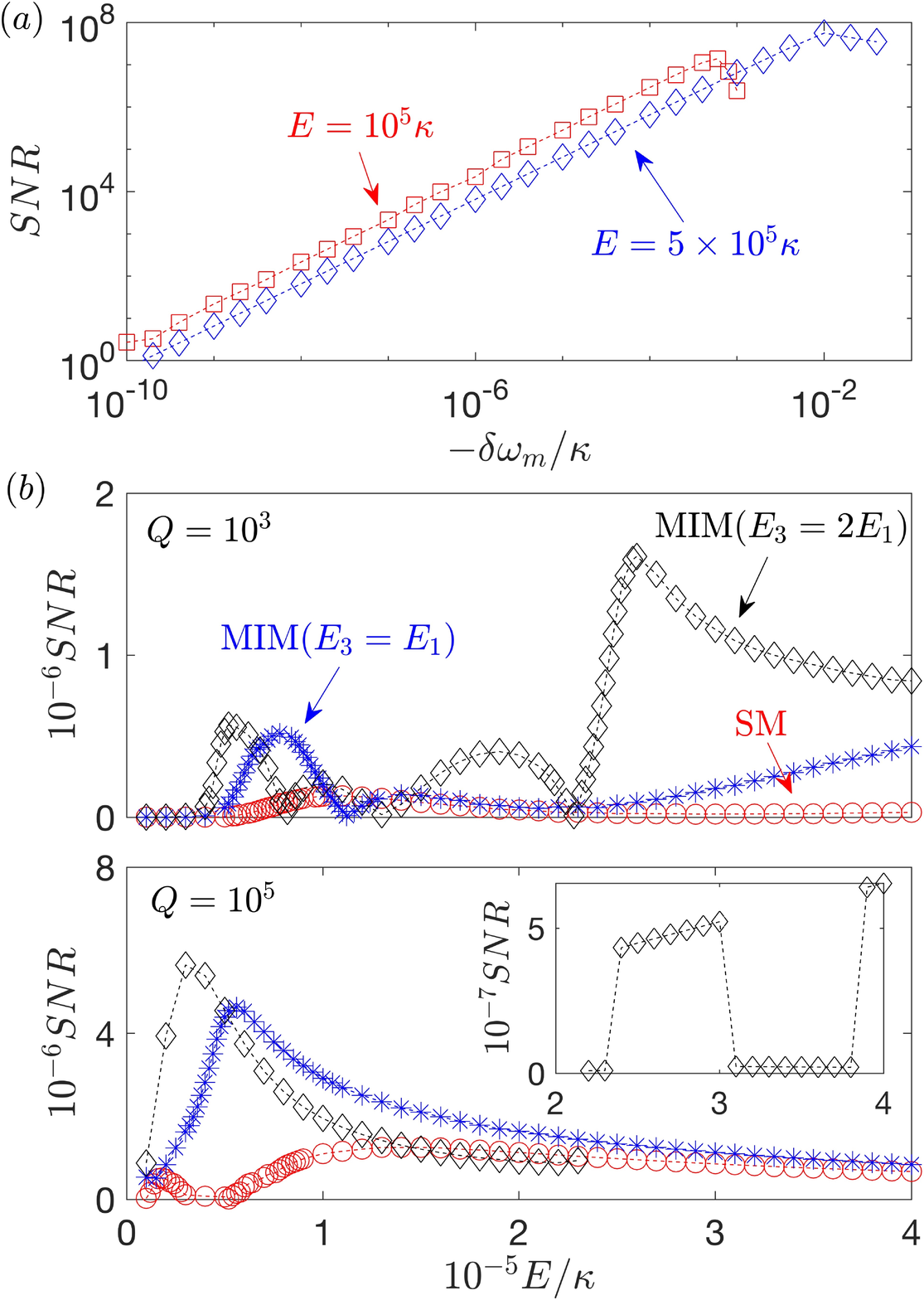}
	\caption{(a) The ranges of the linear scaling between the SNR and frequency shift $\delta \omega_m$ for two different drive amplitudes. The SNR achieved with $E=10^5\kappa$ is larger but has a shorter scaling range.
		(b) The relations between the SNR and the driving field amplitude for an SM system and an MIM system (under two different combinations of driving fields), given the fixed mechanical frequency shift $\delta\omega_m=-10^{-4}\kappa$ induced by a tiny mass added to the mechanical membrane.
		The fixed system parameters are the same as those in Fig. \ref{fig2}.}  \label{fig4}
\end{figure}
	
	The linear scaling relations in Fig. \ref{fig3} are highly useful to the mass sensing. They are originated from the linear relation $\delta A_m=const\cdot \delta\omega_m$ in Fig. \ref{dam}(b), when 
	the frequency shift $\delta \omega_m$ is within the indicated range there. Because the Bessel functions in the expressions of the sideband amplitudes in Eq. (\ref{field}) have the variable 
	$g_mA'_m/\omega_m$, where $A'_m=A_m+\delta A_m$, all of the functions defined for the figures-of-merit (as the differences between after and before adding the measured mass $\delta m$) are proportional to the mechanical amplitude variation $\delta A_m$ from the Taylor expansions to the first order of $\delta \omega_m$. Meanwhile, the observed mechanical oscillation frequency is locked to $\omega_m$ (the quantity in the denominator of $g_mA_m/\omega_m$) by two drives tones satisfying Eq. (\ref{match}). In the regimes where the linear relation $\delta A_m=const\cdot \delta\omega_m$ is valid, a tiny mass $\delta m$, which is mapped to the shift $\delta \omega_m$ of the theoretical mechanical frequency through Eq. (1), can be well determined by the measured sideband intensity change $\delta n^{(2)}_{c,out}$. Then, the range of the linear scaling indicates another important figure-of-merit for such systems of mass sensing. In Fig. \ref{fig4}(a), we provide two examples of the linear scaling range for an MIM system under two different pump power (the three drives tones are assumed to have the same amplitude $E$). It is seen that for the higher drive amplitude $E=5\times 10^5\kappa$, the operation range spans from $\delta\omega_m=-10^{-10}\kappa$ to $-10^{-2}\kappa$. With the lower drive amplitude $E=10^5\kappa$, the linear scaling relation will be violated when the frequency shift $|\delta\omega_m|$ is larger than $6\times10^{-4}\kappa$. The violation of the linear relation is due to the loss of frequency locking after the deviation $\delta \omega_m$ from the condition in Eq. (\ref{match}) becomes too large. Nevertheless, the operation range for this type of mass sensor can theoretically cover from $6$ to $8$ orders of the measured mass.

	\begin{figure*}[t]
		\centering\includegraphics[width=18cm]{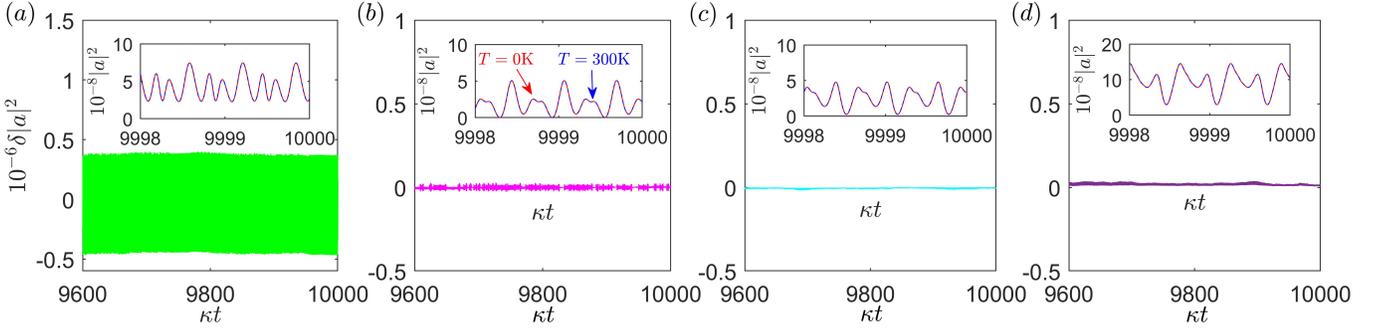}
		\caption{Cavity field intensity change between $T=0$ K and $T=300$ K, for a COMS driven by a single-tone field in (a), an SM system driven a two-tone field satisfying Eq. (\ref{match}) in (b), and an 
			MIM system with $E_3=E_1$ in (c) and $E_3=2E_1$ in (d). The correspondingly evolved cavity intensities are in the insets. The fixed parameters are the same as those in Fig. \ref{fig2} and the pump tone amplitudes are the same as those in Fig. \ref{fig4}.}  \label{fig5}
	\end{figure*}
	
	\begin{figure}[b]
		\centering\includegraphics[width=8.5cm]{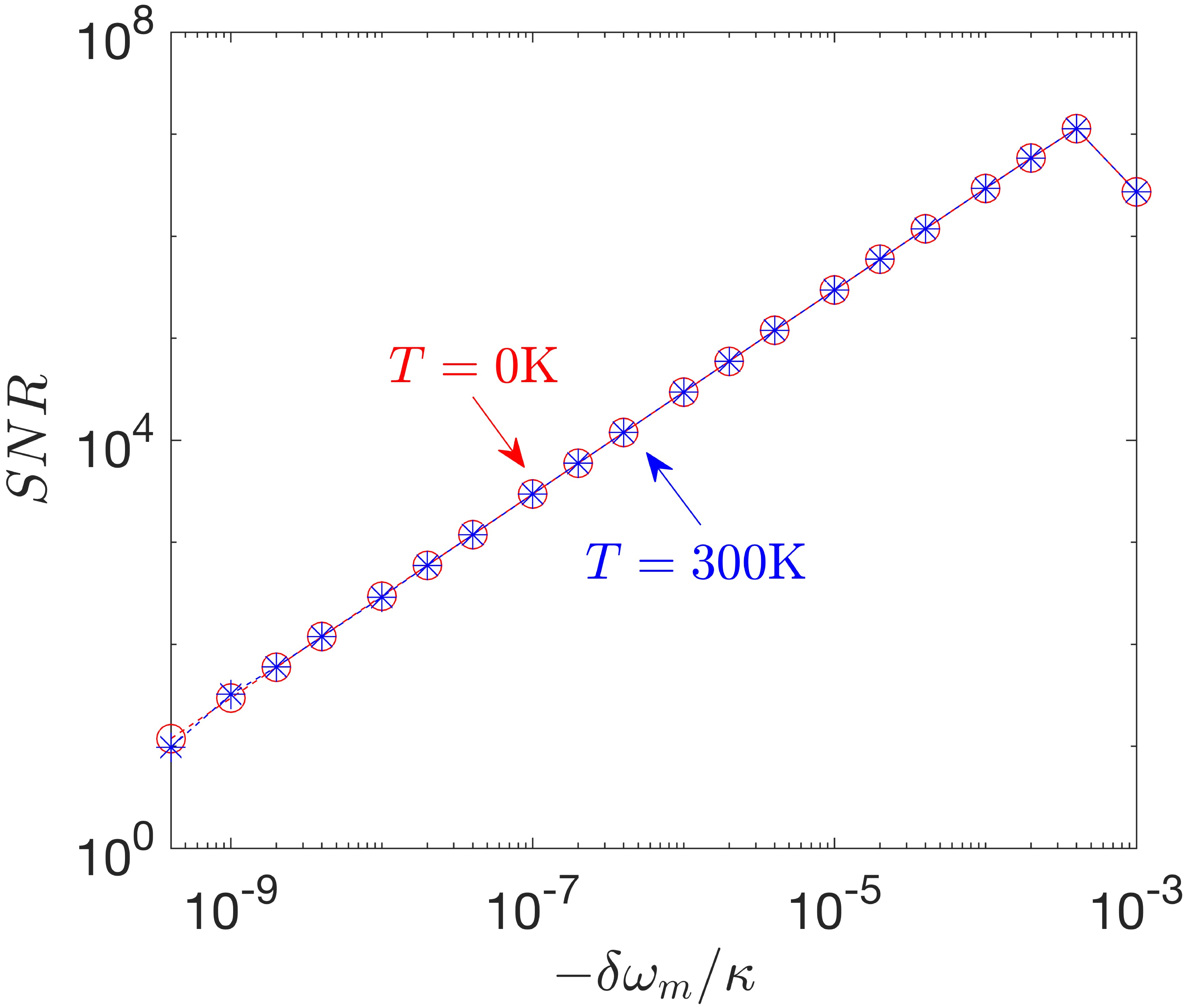}
		\caption{The linear scaling relations between the SNR and $\delta\omega_m$, obtained for an MIM under three drive tones with $E=10^5\kappa$. In the same operation range $\delta\omega_m=-10^{-3}\sim-10^{-9}\kappa$, the SNR at two temperatures are almost the same. The system parameters are the same as those in Fig. \ref{fig2}.}  \label{fig6}
	\end{figure}
	
	\begin{table}[h]
		\centering
		\begin{tabular}{|c|c|c|c|}
			\hline
			&  SM  & MIM ($E_3=E_1$) & MIM ($E_3=2E_1$) \\
			\hline
			$Q=10^3$ & $1.57\times10^{-10}$ & $3.86\times10^{-11}$ & $1.24\times10^{-11}$\\
			\hline
			$Q=10^5$ & $1.60\times10^{-11}$ & $4.32\times10^{-12}$ & $3.54\times10^{-12}$\\
			\hline
		\end{tabular}
		\caption{A comparison of the values of the resolution $\delta m/m$ for two different types of systems and two different mechanical quality factors $Q$, based on the system parameters in Fig. \ref{fig2} and Fig. \ref{fig22}.}
		\label{table1}
		\vspace{-0.3cm}
	\end{table}

We list in Table I the resolutions of the systems for the minimum $\delta m$ measured with either SM system or MIM system, and they are obtained with the limit SNR$=1$ for the measurements. The system of MIM can achieve a better resolution by one order. Its even more meaningful advantage is a much high detection efficiency in the measurement of any fixed tiny mass $\delta m$. The simulated SNR with two different mechanical quality factors $Q$ are very different and also changes with the applied pump field power; see Fig. \ref{fig4}(b). At the optimum drive amplitudes, the SNR can be greatly magnified by the power of the third pump on the MIM system and are significantly higher than those that can be possibly realized by the SM system.

	%\begin{table}[h]
	%\centering
	%\label{table1}
	%\begin{tabular}{|c|c|c|c|c|c|}
	%\hline
	%$\omega_m/\kappa$ & $Q$ & $|\delta\omega_m|/\omega_m$ \cite{qlin1} & $|\delta\omega_m|/\gamma_m$ \cite{qlin1} & $|\delta\omega_m|/\omega_m$ & $|\delta\omega_m|/\gamma_m$\\
	% \hline
	%10 & $10^3$ & $10^{-9}$ & $10^{-6}$ & $5.7\times10^{-11}$ & $5.7\times10^{-8}$   \\
	% \hline
	% 10 & $10^5$ & $10^{-10}$ & $10^{-7}$ & $8.8\times10^{-12}$ & $8.8\times10^{-9}$   \\
	%\hline
	%100 & $10^5$ & $10^{-12}$ & $10^{-7}$ & $1.1\times10^{-14}$ & $1.1\times10^{-9}$  \\
	% \hline
	%\end{tabular}

	%\caption{Mass sensing performance with different mechanical frequencies $\omega_m$ and quality factors $Q$. The drive amplitude is set to be $3\times10^5 \kappa$, and the system parameters as those used in Fig. \ref{fig2}. The sensitivity with the present scheme is one or two-order higher than the former one.}
	%\end{table}

	\section{Operations at room temperature}
	\label{sec4}
	
	A thermal environment is inevitable in many situations that require mass sensing. Due to the impairment by thermal noise, however, most experiments of mass sensing through a direct detection of mechanical frequency shift had to be performed at low temperatures. Unlike the other schemes, our scheme is realized by an oscillation locking mechanism, and this mechanism can also suppress the effect of thermal perturbations during measurement, so that the requirements on the operation environment can be greatly relaxed. In more details, a locked mechanical oscillation by two drive tones satisfying Eq. (\ref{match}) exhibits a robustness against external perturbations: once the dynamical processes have evolved to the locked oscillations, their amplitudes can be hardly modified by a sufficiently strong perturbation, and a noise perturbation can only affect the systems during the transient periods and determine to which locked oscillation orbit the systems will be stabilized \cite{level1}. Such robustness against external perturbations can be also applied to the precise measurement of other physical quantities \cite{f1,f2}.
	
	We simulate the dynamical processes in thermal environment by constructing a thermal noise drive term $\sqrt{2\gamma_m}\xi_m(t)$, which is generated by a random function with Matlab, and the thermal phonon number is set to be $n_{th}=6.244\times10^5$ for the mechanical frequency $\omega_m=10$ MHz and at the environment temperature $T=300$ K. Since what is used in the mass sensing is a cavity field sideband, one needs to know how this noise drive will affect the cavity field. The most obvious quantity is the modification $\delta |a|^2$ of the cavity field intensity or cavity field photon number by the noise drive term $\sqrt{2\gamma_m}\xi_m(t)$. In Fig. \ref{fig5} we present such modifications in a number of different situations. When driven by a single-tone drive, the noise induced modification is obvious as seen from Fig. \ref{fig5}(a), being about $3$ orders lower than the average cavity field intensity. However, once the mechanical oscillation is locked by a two-tone satisfying Eq. (\ref{match}), the deviation $\delta |a|^2$ from the cavity field intensity evolved at $T=0$ K will be greatly suppressed; see Figs. \ref{fig5}(a), \ref{fig5}(b) and \ref{fig5}(c), where $\delta |a|^2$ are about $4$ orders or $5$ orders lower than the average field intensity $|a|^2$. This fact indicates the effect of thermal noise can become insignificant to the concerned mass sensing. The third pump field acting on a MIM system can enhance the modification $\delta |a|^2$ by thermal noise, so there is a restriction on the used intensity of this driving field in practice.
	
	In Fig. \ref{fig6} we present the simulated SNR with the modified frequency $\delta \omega_m$ at both $T=0$ K and $T=300$ K. We find that the linear scaling ranges and SNR values are the same at both temperatures, implying that the sensing operation will not be affected at the room temperature. Such scaling relations are also the same to the modified output sideband intensities $\delta n^{(2)}_{c,out}$, which are directly used to measure the added mass $\delta m$. A better mechanical quality factor $Q$ or a smaller mechanical damping rate $\gamma_m$ will improve the performance at the room temperature further.

	\section{Discussions and conclusion}
	\label{conclusion}
	
	The setup for the described mass sensing can be developed from a reported experimental SM system \cite{sm1}. In that experiment mechanical oscillations with their amplitudes in the order of $1$ nm are excited with the pump power at the level of $10$ mW (the pumps are of single tone), and the scenario is compatible with our mass sensing operations after the system is also possibly modified to a MIM one and the pump field is replaced by multiple tones. A crucial point for our mass sensing is to enter a locked mechanical oscillation, which can be realized with the system parameters meeting the condition $g_mE/\omega_m\sim 0.1\kappa$. Optimizing the ratio $g_m/\omega_m$ in the fabrication of a COMS is, therefore, important to the reduction of pump power in operations. 
	
	One of the important results of the current work is the existence of a small scaling window between the amplitude modification and the theoretical frequency shift of the observed optomechanical oscillations, when the exact pump-tone condition in Eq. (\ref{match}) is slightly violated. Such scaling window illustrated in Fig. \ref{dam}(b), which was not applied in a previous work of the same category \cite{qlin1}, is the foundation for our mass sensing, and the associated figures-of-merits (the high-order sidebands' modifications and the SNR) for the operations also follow their corresponding linear scaling relations. The scaling window will be the largest if two drive tones have their difference to be tuned exactly to the original theoretical frequency $\omega_m=\sqrt{k/m}$. With the currently available technologies \cite{AOM, SS}, the difference of two drives tones can be tuned close to this mechanical frequency in the orders of $1$ MHz $\sim$ $100$ MHz, but an error in matching the condition in Eq. (\ref{match}) can reduce the scaling range for operation. In the measurement of a tiny mass $\delta m$ corresponding to the $\delta \omega_m$ on the left-hand side of Fig. \ref{dam}(b), however, an error is allowed to result in a shorter scaling range. 
	
	In conclusion, we have detailed a mass sensing approach based on the nonlinear dynamics of MIM COMS driven by three tones of pumping laser fields. With two of the drive tones tuned close to the frequency condition in Eq. (\ref{match}), the frequency of a stabilized mechanical oscillation of the COMS will be locked to the difference of these two tones. If an added unknown mass to the mechanical membrane is sufficiently tiny, the observed mechanical oscillation frequency will not be changed, but the cavity field sidebands will be slightly modified in their amplitudes. Within a detection window, the changes of sideband amplitudes are linearly proportional to the unknown mass, so that the measurement of the mass can be realized by precisely reading the modified field sideband amplitude. More importantly, the application of the third pump tone can magnify the sideband's modification. Compared with the scheme based on the SM systems \cite{qlin1}, the MIM configuration significantly improves the mass sensing efficiency [see Fig. \ref{fig4}(b)], in addition to the enhancement of the detection sensitivity $\delta m/m$ by one order. Due to the robustness of the locked mechanical oscillation against noisy perturbations, the measurement can be well perform at room temperature. These features can greatly relax the restriction on the realization of practical ultrahigh resolution mass sensing. As an example of important application, the current study elucidates the necessity for exploring the COMS with two cavity fields or two mechanical modes \cite{sm1,mi1,mi2,mi3,mi4,mi5, mi6}, as well as those of even more system modes \cite{extra0,extra1,extra2,extra3,extra4,extra4b,extra5,extra6,extra7,extra8}. The further development of technologies may lead to the real usage of these more sophisticated systems.  
	
	\begin{acknowledgments}
		This work was supported by National Natural Science Foundation of China (Grant No. 12374348), Natural Science Foundation of Fujian Province (Grant No. 2024J01078), and ANID Fondecyt
		Regular (Grant No. 1221250). 
	\end{acknowledgments}

\end{document}